# Can Density Matrix Embedding Theory with the Complete Activate Space Self-Consistent Field Solver Describe Single and Double Bond Breaking in Molecular Systems?

*Hung Q. Pham, Varinia Bernales, and Laura Gagliardi\**

Department of Chemistry, Chemical Theory Center, and Minnesota Supercomputing Institute, University of Minnesota, Minneapolis, Minnesota 55455-0431, United States

**ABSTRACT**

Density matrix embedding theory (DMET) [*Phys. Rev. Lett.* **2012**, *109*, 186404] has been demonstrated as an efficient wave-function-based embedding method to treat extended systems. Despite its success in many quantum lattice models, the extension of DMET to real chemical systems has been tested only on selected cases. Herein, we introduce the use of the complete active space self-consistent field (CASSCF) method as a correlated impurity solver for DMET, leading to a method called CAS-DMET. We test its performance in describing the dissociation of a H–H single bond in a $H_{10}$ ring model system and an N=N double bond in azomethane ($CH_3$–N=N–$CH_3$) and pentyldiazene ($CH_3(CH_2)_4$–N=NH). We find that the performance of CAS-DMET is comparable to CASSCF with different active space choices when single-embedding DMET corresponding to only one embedding problem for the system is used. When multiple embedding problems are used for the system, the CAS-DMET is in a good agreement with CASSCF for the geometries around the equilibrium, but not in equal agreement at bond dissociation.

## I. INTRODUCTION

Quantum-chemical methods for macromolecules and extended systems play a crucial role in chemistry, physics, and material sciences.[1-7] For these systems, Kohn-Sham density functional theory[8-9] (KS-DFT) is routinely used due to its affordable computational cost. However, the lack of knowledge about the exact functional prevents the application of KS-DFT for critical systems where the mean-field approximation fails to describe the correct physics.[10-13] Wave function theory, on the other hand, can be systematically improved via hierarchical approximations.

Conventionally, the exact solution for a quantum system can be computed by means of the full configuration interaction (FCI) expansion.[14-15] This method, in practice, is too expensive to be useful and FCI solutions are only available for small systems with small basis sets. For small and medium size molecular systems with a single reference electronic structure, a chemical accuracy of 1 kcal mol$^{-1}$ can be reached using coupled-cluster single double excitations with perturbed triples theory (CCSD(T)).[16-17] For critical systems where strong electron correlation is significant, the traditional complete-active space self-consistent field theory[18-20] (CASSCF) and its derivatives like restricted active space SCF[14] (RASSCF), generalized active space SCF[21-23] (GASSCF), or occupation-restricted-multiple-active-space SCF[24-25] (ORMAS) can be employed to treat static correlation, while dynamic correlation can be recovered using post SCF treatments, like second-order perturbation theory[26-29] (CASPT2, RASPT2, ORMAR-PT), or multiconfigurational pair density functional theory[21, 30] (MC-PDFT), on top of these wave functions. Furthermore, other multiconfigurational ansatzes inspired by tensor network theory,[31-32] e.g. density matrix renormalization group[33-35] (DMRG), have been successful for medium-sized chemical systems, especially for those with a one-dimensional topology. Despite many quantum chemical methods, the accurate treatment of strongly correlated extended systems is a challenge due to the exponential scaling of wave function methods with respect to the system size, usually defined as the exponential wall of quantum chemistry.[36]

Recently, density matrix embedding theory[37-38] (DMET) has emerged as an efficient wave-function-based embedding approach to treat strongly correlated systems. The success of DMET lies in the fact that the total quantum system can be exactly partitioned into smaller subsystems or fragments that can be treated by any of the high-level quantum chemical solvers previously mentioned. In this way, one only has to calculate a set of subsystems with high accuracy instead of dealing with the intractable total system at the same level of theory. Mapping the system into an impurity embedded in an effective environment is the key idea behind dynamic mean-field theory[39-41] (DMFT), a successful computational methodology for strongly correlated materials.[42] Instead of formulating the quantum impurity problem using many body Green's function ansatz,[43] DMET has its root in the frequency-independent local density matrix, which makes the ansatz simpler and more efficient with similar accuracy in comparison to DMFT.[38] DMET has demonstrated its versatility and success in providing the accurate solutions for one- and two-dimensional and honeycomb Hubbard models,[38, 44-45] strongly correlated spin systems,[46-47] and some chemical systems.[37, 48-49] Since its advent in 2012, many derivatives of DMET have been proposed. These include density embedding theory[44] (DET) with a broken symmetric mean-field bath and a diagonal-only matching scheme for density matrix, DMET with an antisymmetrized geminal power bath[50] instead of a mean-field bath, bootstrap embedding DMET[51-52] with significant improvement of the convergence rate by matching the wave function at the edges and centers of the different fragments, and very recently the cluster or block-product DMET[46-47] for quantum spin systems where the exact embedding bath states are replaced by a set of block-product states. A variety of correlated solvers have been employed in DMET calculation, including FCI,[38, 44, 48] DMRG,[37] and auxiliary-field quantum Monte Carlo.[53]

Despite the success of DMET in many lattice models, its application to real chemical systems has not been extensively explored. To the best of our knowledge, only coupled-cluster with double (CCD) and single and double (CCSD) excitations solvers have been employed for a few strongly correlated molecular and periodic systems,[37, 48-49] and for model systems like a hydrogen ring or hydrogen rectangular lattice where the FCI solutions are tractable under small basis set.[48] In this work, we introduce the use of CASSCF as a correlated impurity solver for DMET, namely

CAS-DMET. We demonstrate the application of CAS-DMET by computing bond dissociation energies for (i) the simultaneous H–H bond dissociation in a $H_{10}$ ring model system; (ii) N=N double bond dissociation in azomethane (CH$_3$–N=N–CH$_3$) and pentyldiazene (CH$_3$(CH$_2$)$_4$–N=NH), two realistic molecular systems. For these systems, different active space selections and energetic comparisons between CAS-DMET and CASSCF are presented. We also visualize and analyze the molecular orbitals for different active spaces choices in DMET to interpret its performance for chemical bond dissociation. The paper is organized as follows. In section II we describe DMET theory and in section III the computational methods employed in this study. In section IV we present our results and their discussion and in section V we offer some conclusions.

## II. THEORY

We briefly summarize the basic equations of density matrix embedding theory before describing the use of complete active space self-consistent field (CASSCF) as the impurity solver. For further details on the DMET formulation, one should read the comprehensive articles.[37, 54] Consider a full quantum system that can be partitioned into a subsystem $A$ of interest, defined as an impurity, or cluster, or fragment, surrounded by an environment $B$. In DMET one embeds the fragment $A$ in a quantum bath that represent the entanglement of the embedded fragment with the environment. The bath has the same number of many-body basis states ($N_f$) as those used to describe $A$. The exact wave function of the total system undergoes a Schmidt decomposition:[55]

$$|\Psi\rangle = \sum_i^{N_f} \lambda_i |\alpha_i\rangle \otimes |\beta_i\rangle \tag{1}$$

where $|\alpha_i\rangle$ and $|\beta_i\rangle$ are the fragment and bath states, respectively; $\lambda_i$ are the singular values of the singular value decomposition (SVD) of the coefficient tensor; $N_f$ is the dimension of the fragment $A$, which is assumed to be smaller in size than the other part of the system. In practice, the Schmidt decomposition for the exact wave function cannot be obtained, and one often uses a mean-field wave function to construct the effective bath states. The Schmidt form of a Hartree-Fock (HF) wave function[56] is given by:

$$|\Psi_{\text{HF}}\rangle = \sum_i^{N_f} \lambda_i |\alpha_i\rangle \otimes |\beta_i\rangle \otimes |core\rangle \quad (2)$$

where $|core\rangle$ represents the core (or pure environment) states, also referred to as core orbitals, which are unentangled to the local fragment states. The pure environment is a similar concept to the core contribution in the frozen-core approximation of the active space ansatz.[57] This decomposition of the HF solution can be obtained using a mean-field one-particle density matrix or a projector onto a set of fragment basis functions.[44, 48] The Schmidt decomposition yields single-particle embedding bases (fragment plus bath states), also known as a DMET active space, that can be used to project the Hamiltonian onto the embedding space. Notice that at this point the dimension of the embedding problem has been considerably reduced compared to the original problem and it is at most twice the number of the basis functions ($2N_f$) that span the fragment $A$. Now, a high-level quantum chemical solver could be used to solve this embedding wave function (note that the core states are excluded in this case).

In general, a chemical system can be partitioned into multiple arbitrary fragments ($A_x$), where each fragment is associated to its own embedding wave function. The total energy $E_{\text{total}}$ of the system can be recovered from the fragment energies $E_x$, i.e. $E_{\text{total}} = \sum_x E_x + E_{\text{nuc}}$ where $E_{\text{nuc}}$ is the nuclear-nuclear repulsion energy and *x* is the number of fragments in which the entire system has been partitioned.[37] The arbitrary partition of a chemical system and the fact that we construct the bath from an approximate wave function necessitates enforcing the total number of electrons calculated by adding up those of the fragments to be equal to the number of electrons of the total system. A variable, known as a global chemical potential ($\mu_{global}$), is added to the one-particle part of the embedding Hamiltonian and is optimized during the DMET procedure. It should be noted that the chemical potential becomes redundant in the DMET formalism for periodic systems or systems with translational symmetry where the number of electron per fragment (a unit cell) is well defined.[49] In addition to the global chemical potential, the one-particle Hamiltonian is augmented by an effective correlation potential $\hat{C}_x$ for each fragment $A_x$, i.e. $\hat{h}' = \hat{h} + \sum_x \hat{C}_x$. The correlation potential is self-consistently varied to minimize the difference between the one-particle reduced density matrix (1-RDM) calculated at the mean-field level and the one calculated by the high-level method, thus improving the bath representation after each self-consistent cycle. Various matching conditions can be applied in order to minimize the difference between the mean field and high-level 1-RDM. One can match the entire embedded 1-RDM or only those elements corresponding to the fragment states. If only the diagonal elements of the embedding 1-RDM are considered during the matching process, this results in density embedding theory (DET).[44] In practical calculations, one can keep the bath states constructed from the HF wave function unchanged, and this is defined as a one-shot DMET calculation. Aside from the multiple-partition way of performing DMET, one can partition the system only one time, defined as a single embedding DMET, and use the following formula to recover the total system energy:

$$E_{total} = E_{CAS\ (\equiv fragment\ +\ bath)} + E_{core} + E_{nuc} \qquad (3)$$

where $E_{CAS}$, $E_{core}$, and $E_{nuc}$ are the electronic energy contribution from the DMET active space constituted by fragment and bath states, core states, and the nuclear-nuclear repulsion energy, respectively.

The fundamental idea behind the CAS-DMET is to introduce a CASSCF active space inside the DMET space (Figure 1). Some natural questions about this method are:

i) How do the results of this procedure differ from a conventional CASSCF calculation on the full system?

ii) How dependent is the method performance on the active space choice?

Concerning i), in the one-shot DMET calculation, the core states are kept frozen and the CASSCF solver rotates only the fragment and bath states in a self-consistent fashion. In a conventional CASSCF calculation on the full system, on the other hand, the entire orbital space (usually HF orbitals) is optimized. Therefore, a one-shot CAS-DMET calculation has fewer electronic degrees of freedom than a CASSCF calculation for the same active space. Hence, the CAS-DMET energy is expected to be higher than that of the CASSCF if the correlation potential is not optimized. We will address ii) in the following.

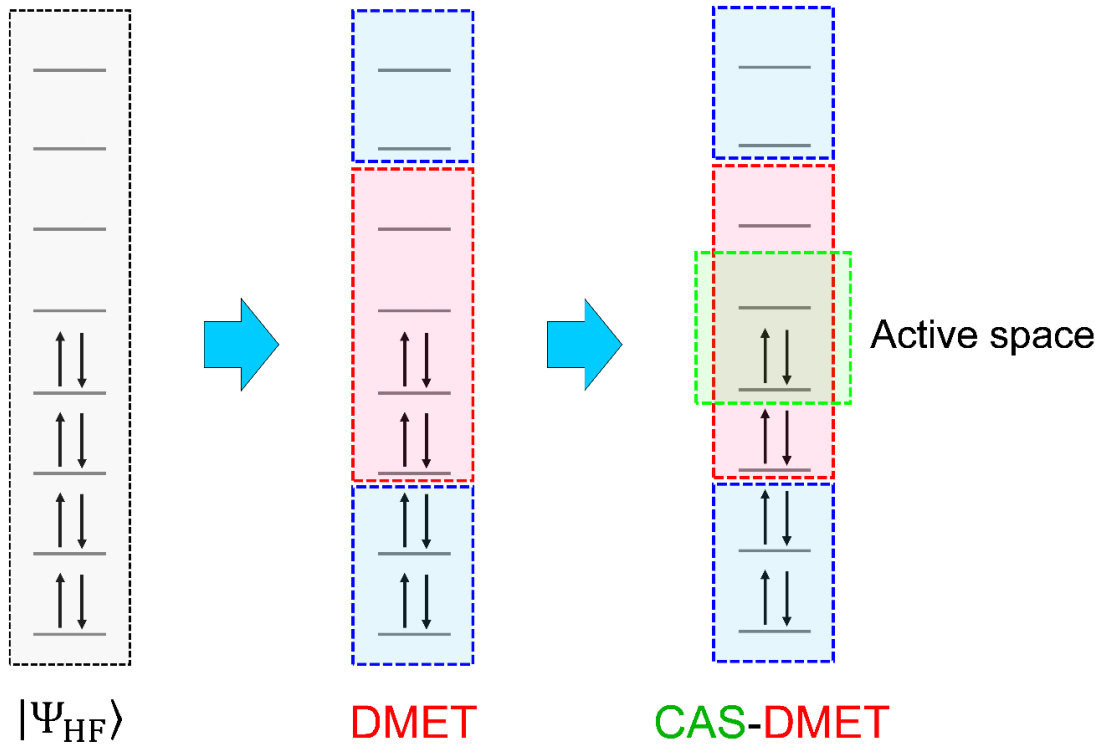

**Figure 1**. CAS-DMET set of orbitals. Starting from a Hartree-Fock wave function (black box), the DMET orbitals are generated, these include: fragment and bath (red box), pure environment or core state (blue box). The CASSCF active space is represented as a green box inside a DMET active space in red.

## III. COMPUTATIONAL METHODS

The CAS-DMET calculations were performed with our locally modified version of QC-DMET.[58-59] As high-level method we used the CASSCF method as implemented in PySCF.[35] The DMET solver solves the embedding Hamiltonians and returns the fragment energies and the CASSCF one-particle density matrix.

The following systems were explored: the dissociation of an H–H single bond in a $H_{10}$ ring model system, and the dissociation of an N=N double bond in azomethane ($CH_3$–N=N–$CH_3$) and pentyldiazene ($CH_3(CH_2)_4$–N=NH).

All geometries were optimized using the B3LYP[37,38] density functional as implemented in the *Gaussian 09* software package,[36] together with the 6-31g(d,p)[60-61] basis set for C, H, and N atoms. The nature of all stationary points for all the structures of interest was verified by analytical computation of vibrational frequencies. Subsequently, single point calculations at different H–H or N=N bond lengths while the other geometric parameters are kept at the equilibrium geometry were performed using CASSCF and CAS-DMET. The bond dissociation energy was computed as the difference between the minimum energy and the energy at large separation between the fragments. In practice, a DMET calculation often starts with an orbital localization in order to generate a set of orthonormal orbitals for the bath construction. The meta-Löwdin orbital localization[36] scheme was used in the bath constructions. The 6-31G basis set was used for all energy calculations. The molecular orbitals from CASSCF and from CAS-DMET are presented in section IV. The orbital transformation in the CAS-DMET space is discussed in Appendix I. Note that for CASSCF and CAS-DMET calculations, an active space of $n$ electrons in $m$ orbitals is denoted as ($n$, $m$).

## IV. RESULTS AND DISCUSSION

### A. Hydrogen ring ($H_{10}$)

In this section, we discuss the performance of one-shot CAS-DMET in computing the symmetric dissociation of a ring of 10 hydrogen atoms. While the DMET calculation using FCI solver has been reported previously,[37, 48] herein we concentrate on comparing the results obtained with CAS-DMET versus those with CASSCF on the total system (Figure 2). The CAS-DMET calculations were performed for different active spaces using two hydrogen atoms as impurity, and correspondingly four $s$ orbitals as impurity orbitals. Using the 6-31G basis set, the total FCI problem corresponds to 10 electrons in 20 orbitals, while the DMET space contains 8 electrons in 8 orbitals, including four fragment and four bath orbitals. Hence, the FCI limit of the embedding problem is (8,8), while that of the total system is (10,20). Considering the symmetry of the system, we just have five identical embedding problems (five subsets of two hydrogen atoms as fragments) that have a smaller Hilbert space compared to the original problem. Figure 1 shows the dissociation energies of $H_{10}$ calculated by CAS-DMET and CASSCF using different active spaces. For the CAS-DMET calculations, we systematically expand the active space size by adding two electrons and two orbitals at a time to a minimal active space of (2,2), i.e. two electrons in a space composed of the highest-occupied molecular orbital (HOMO) and the lowest-unoccupied molecular orbital (LUMO). This results in three active spaces: (2,2), (4,4), (8,8), respectively. For the CASSCF calculations, we consider five active spaces: (4,4), (8,8), (10,10), (10,12), (10,14) in which we choose an equal number of the occupied and unoccupied orbitals around HOMO and LUMO for the first three active spaces, and then add more unoccupied orbitals for the last two active spaces. Our results show that CAS-DMET(4,4) predicts the same dissociation curve as CAS-DMET(8,8), which reaches the FCI limit in the embedding space. A comparison to the CASSCF energies shows that the CAS-DMET(8,8) and CASSCF(10,14) energies near the equilibrium geometry agree within 1 $mE_h$ (Figure 1b). However, for distances larger than 1.5 Å, the CAS-DMET energies with both the (4,4) and (8,8) active spaces are lower than those of CASSCF with the (10,10), (10,12), and (10,14) active space. The CASSCF energies with the (10,12), and (10,14) active spaces are asymptotically equal to the correct value of 0.5 $E_h$ while both CAS-DMET(4,4) and CAS-DMET(8,8) predict an inaccurate infinite dissociation energy of ca. 0.508 $E_h$. This discrepancy can be attributed to the approximate bath states used in the one-shot calculation. This could be improved by interactively optimizing the correlation potential. However, in order to analyze the orbitals, we did not employ the self-consistent procedure since the orbitals would change during the interactive cycles.

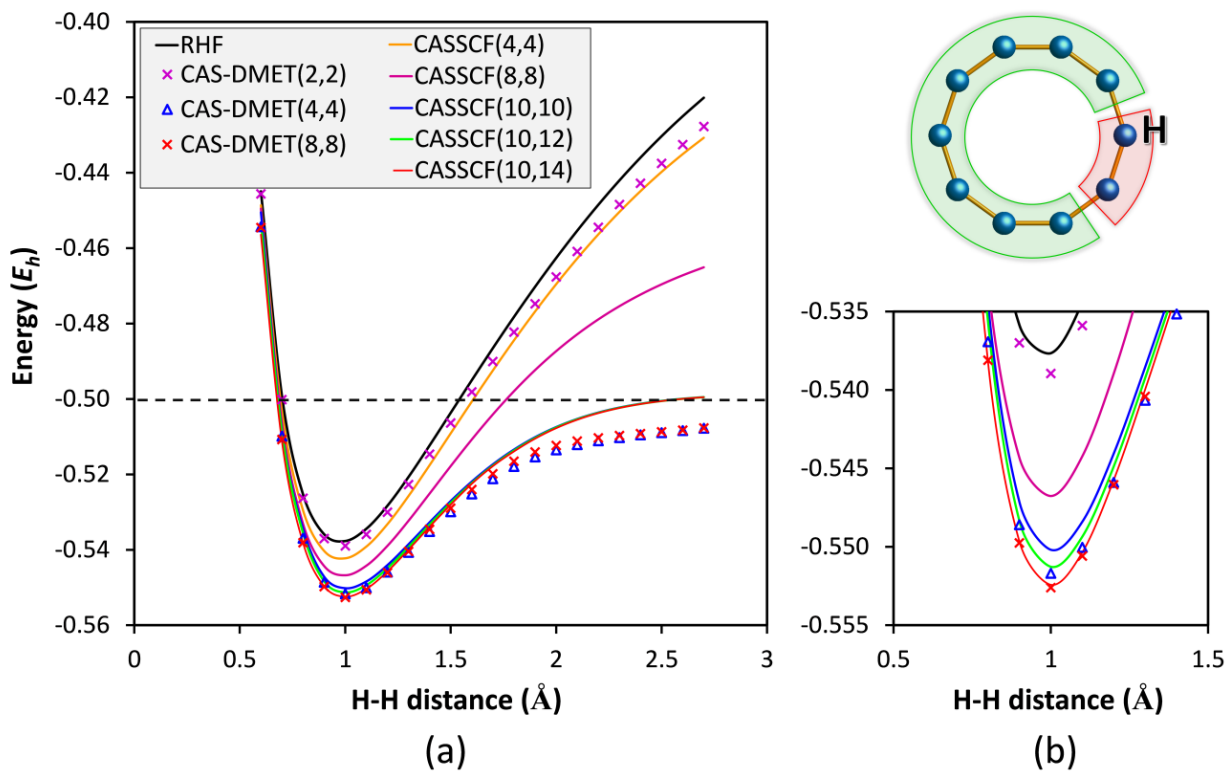

**Figure 2**. Simultaneous dissociation of $H_{10}$ ring energies (per atom) calculated by using restricted Hartree-Fock (RHF), CAS-DMET, and CASSCF. Plot (a) shows results from the equilibrium region to dissociation, while (b) shows a detailed plot for the 0.50–1.50 Å region. The CAS-DMET was performed using two hydrogen atoms (red box), which correspond to four *s* orbitals as a fragment as shown in this figure (top right). The dashed line indicates the correct dissociation energy of 0.5 $E_h$.

In Figure 3, the $H_{10}$ ring DMET orbitals are presented for the case with four fragment and four bath orbitals, in addition to the 12 core states that belong to the pure environment. We observe that the fragment orbitals are localized on the two hydrogen atoms taken as fragment states, while the bath orbitals are more delocalized over the rest of the system, as expected. This is consistent with the fact that the bath orbitals represent the entanglement between the environment and the local fragment. For more details on the construction of these orbitals, we refer the reader to the Appendix section. Figure 4 shows the CAS-DMET orbitals that are linear combinations of the fragment and bath orbitals (shown in Figure 3).

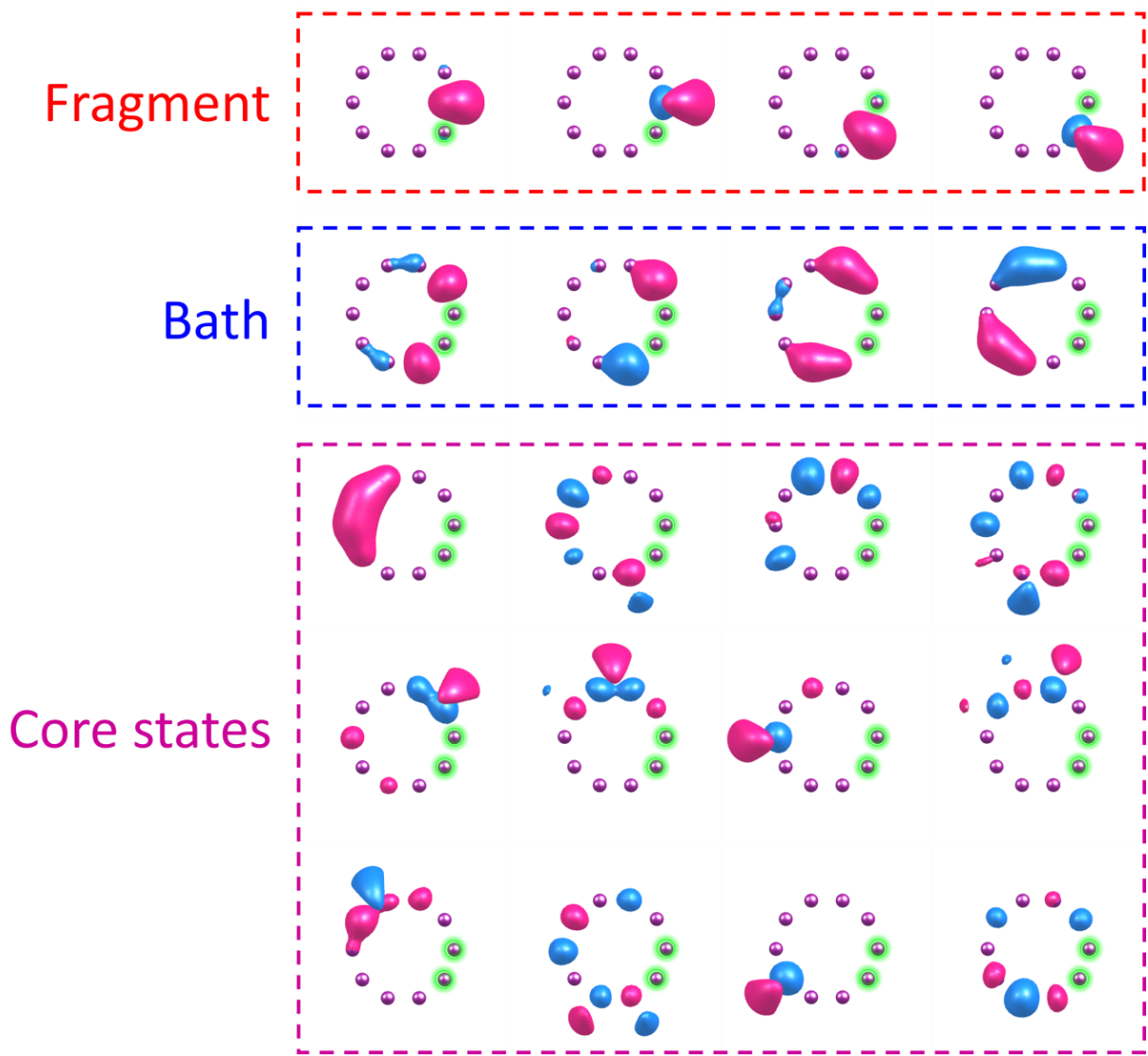

**Figure 3**. The fragment, bath, and core or pure environment orbitals of the $H_{10}$ ring at a separation of 1.0 Å. The fragment atoms are highlighted by green circles.

Figure 4 shows the comparison between the natural orbitals of the CAS-DMET(4,4) and CASSCF(10,10) calculation at an H–H separation of 1.0 Å. These two active spaces are the minimum ones within each method to recover static correlation as demonstrated in the energy curves (Figure 2.b). While the CASSCF orbital are quite delocalized, the CAS-DMET orbitals are localized over four hydrogen atoms (two fragments and two neighbor atoms).

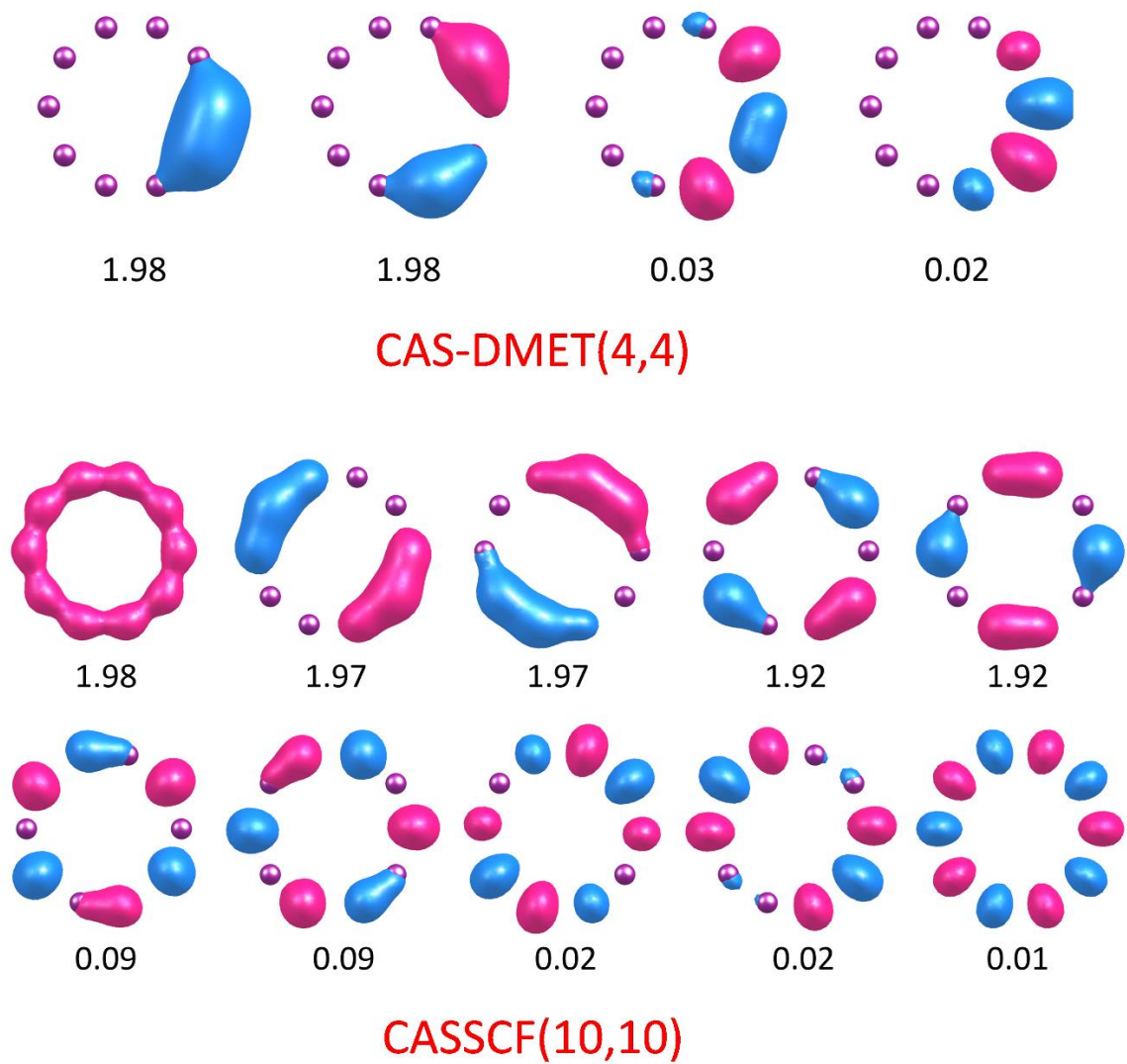

**Figure 4**. CAS-DMET(4,4) and CASSCF(10,10) orbitals and their occupation numbers for $H_{10}$ ring at a separation of 1.0 Å.

**B. Azomethane ($CH_3$–N=N–$CH_3$)**

The purpose of this study is to analyze the similarity between CAS-DMET and CASSCF for a double bond dissociation that occurs in a well-defined fragment of a molecular system. We first present one-shot CAS-DMET results for the dissociation of an N=N double bond in azomethane (Figure 5). We used a multiple partition scheme where we divide the system into three fragments, namely two methyls (CH$_3$–) and one dinitrogen (–N=N–) fragment. The embedding problem of the –N=N– fragment (shown in a red box in Figure 5) was solved by the CASSCF solver, while the other two CH$_3$– fragments (shown in a blue box in Figure 5) were treated at the restricted HF (RHF) level. We considered two different ways of representing the bath orbitals: (1) a full bath, which takes all the entangled orbitals as bath states and (2) a truncated bath, where only the most entangled orbital per bond are used as bath states. For breaking the –N=N– double bond, we hypothesize that the two most entangled orbitals are sufficient to represent the bath states following the argument in ref 62.[62]

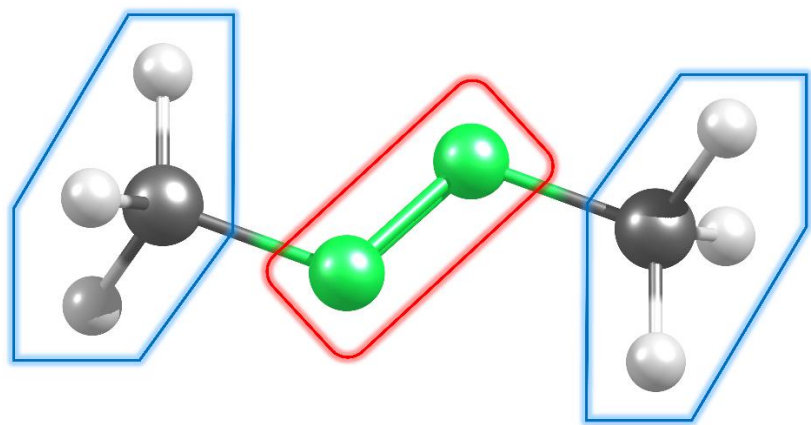

**Figure 5**. Multiple partition in azomethane. –N=N– fragment (red box) is computed by CASSCF solver while the CH$_3$– fragments (blue boxes) are computed by RHF solver. Color code: white, hydrogen; grey, carbon; green, nitrogen.

Since we are breaking a double bond an active space of four electrons in four orbitals is sufficient to describe the static correlation of the system. Figure 6 shows the dissociation energies obtained by CAS-DMET using the full and truncated baths. Not surprisingly, both CAS-DMET(4,4) and CASSCF(4,4) are able to qualitatively describe the dissociation behavior compared to a larger active space of 8 electrons in 8 orbitals. In general, there is good agreement between CAS-DMET and CASSCF energies near the equilibrium bond length (1.0–2.0 Å) for different active spaces and bath schemes (Figure 6). However, for N–N distances > 2.0 Å, the CAS-DMET with the full bath scheme poorly reproduces the CASSCF energies. The truncated bath CAS-DMET performs qualitatively better than the full bath CAS-DMET even though it does not reproduce the total CASSCF energies and the energies are shifted to match the equilibrium energy when compared to the CASSCF energies (see Table S3 for the absolute energies). To understand such a poor performance of the full bath scheme, we further analyzed the orbitals in the active space of the various calculations.

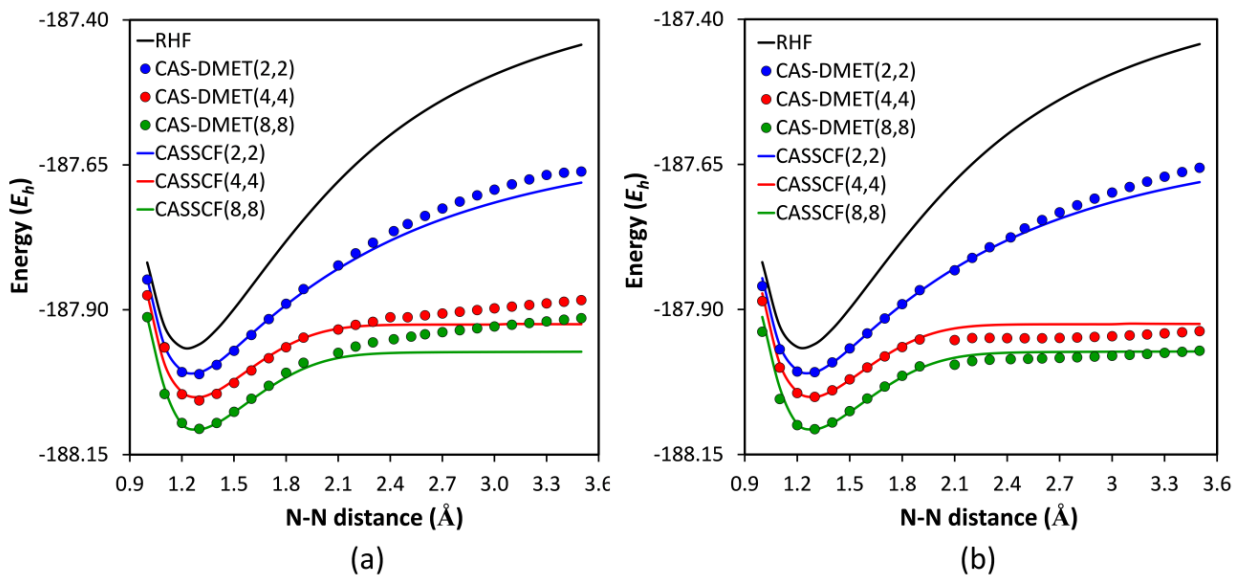

**Figure 6**. CAS-DMET and CASSCF calculations on azomethane: the full bath (a) and the truncated bath (b). Note that the truncated bath energies were shifted to match the minimum energies between CAS-DMET and CASSCF.

Figure 7 shows that at the equilibrium distance of 1.3 Å the CAS-DMET and CASSCF orbitals are similar for both the full and truncated bath. The same observation is valid for the occupation numbers. However, at 3.5 Å, the full bath CAS-DMET orbitals look quite different from the CASSCF ones, while the truncated bath CAS-DMET orbitals are more similar to the CASSCF orbitals, together with their occupation numbers. In particular, all four orbitals from the full-bath CAS-DMET have symmetric shapes while there are two asymmetric orbitals among four orbitals of the truncated CAS-DMET or CASSCF. This is in line with the observation that the truncated scheme gives better agreement with CASSCF in terms of energy than the full scheme in dissociating the N=N double bond.

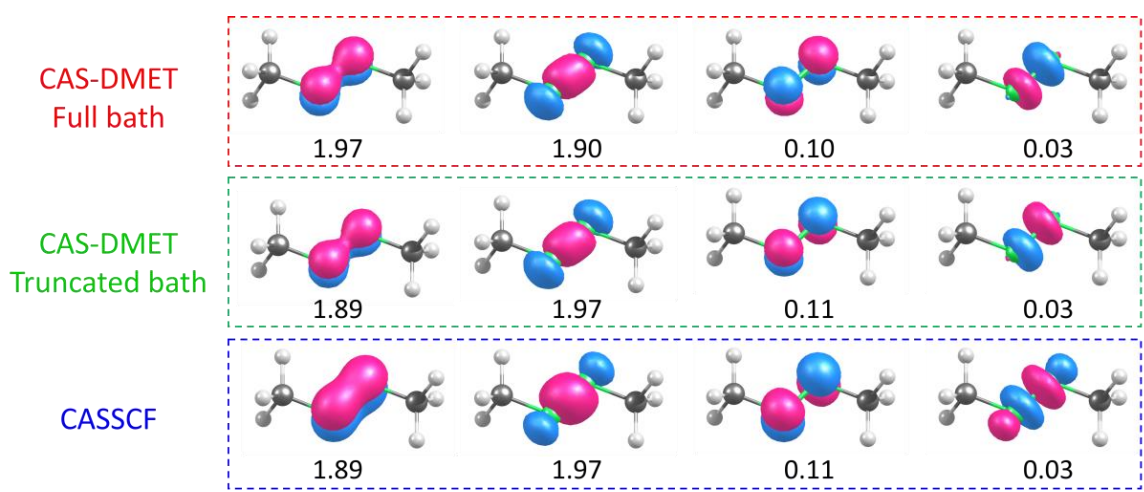

(a)

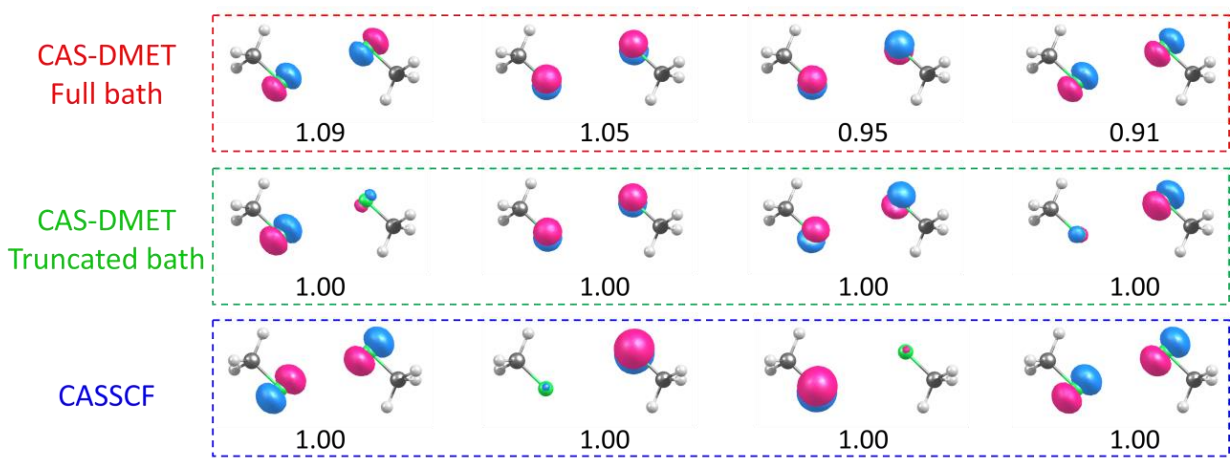

(b)

**Figure 7**. CAS-DMET(4,4) and CASSCF(4,4) natural orbitals and occupation numbers of azomethane at a separation of 1.3 Å (a) and 3.5 Å (b). Full CAS-DMET bath orbitals are enclosed in a red box, truncated CAS-DMET bath orbitals in green, and the CASSCF orbitals in blue. Color code: white, hydrogen; grey, carbon; green, nitrogen.

The fragment and bath orbitals of azomethane are reported in Figures SI1 and SI2 for two N-N distances, to gain insight into what triggers the performance of the truncated bath CAS-DMET. In the full-bath calculation, 18 fragment orbitals and 16 bath orbitals are chosen to form the embedding space based on the degree of entanglement. In the truncated-bath calculation, only the two most entangled bath orbitals are kept in the space. We note that the bath orbitals are approximately constructed using a HF wave function. The fact that eliminating some HF bath orbitals improves the dissociation energy with regard to the CASSCF energy indicates that the truncated bath scheme may be helpful in canceling the error caused by the approximate bath used in DMET.

## C. Pentyldiazene (CH$_3$(CH$_2$)$_4$–N=NH)

In this section, we combine the single embedding scheme with the CAS-DMET solver by computing the N=N dissociation energy in CH$_3$(CH$_2$)$_4$–N=NH. In this case, the environment is larger than in azomethane and we can thus compare the CAS-DMET performance for similar environments, but of different sizes. The N=NH unit is considered as a fragment, and the rest of the molecule is the environment (Figure 8). The total energy of the system is computed using eq. (3). We present both one-shot and self-consistent CAS-DMET results, the latter being denoted as scCAS-DMET from now on.

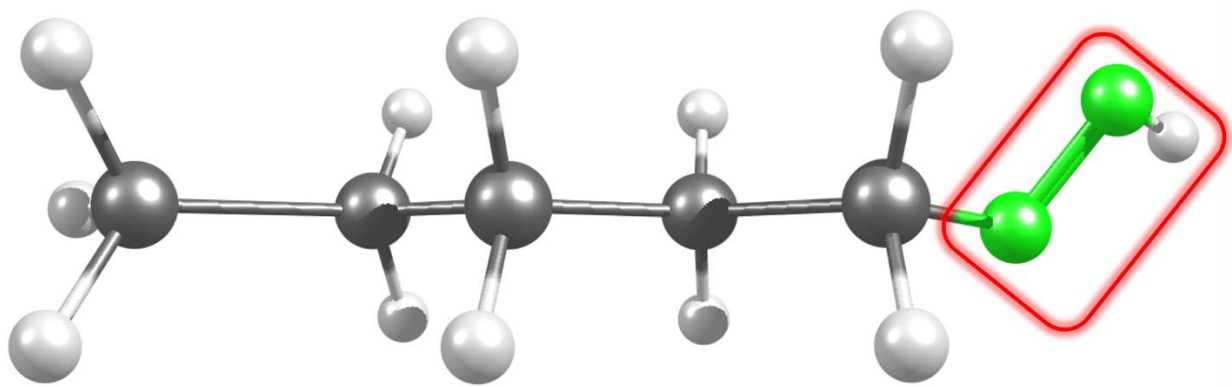

**Figure 8**. Partition of pentyldiazene (CH$_3$(CH$_2$)$_4$–N=NH) molecule in CAS-DMET. The fragment of interest is enclosed in the red box. Color code: white, hydrogen; grey, carbon; green, nitrogen.

There is an excellent agreement between the single embedding CAS-DMET and CASSCF in describing the N=N dissociation energy (Figure 9), even better than in the azomethane case, because the system is not partitioned into multiple fragments that correspond to multiple embedding problems like for azomethane. This single embedding fashion of DMET is similar to the conventional CASSCF since both use only one active space (or the embedding space for DMET) for the system. The self-consistent optimization of the correlation potential, scCAS-DMET, does not change significantly the total energy of the system, while being computationally more demanding.

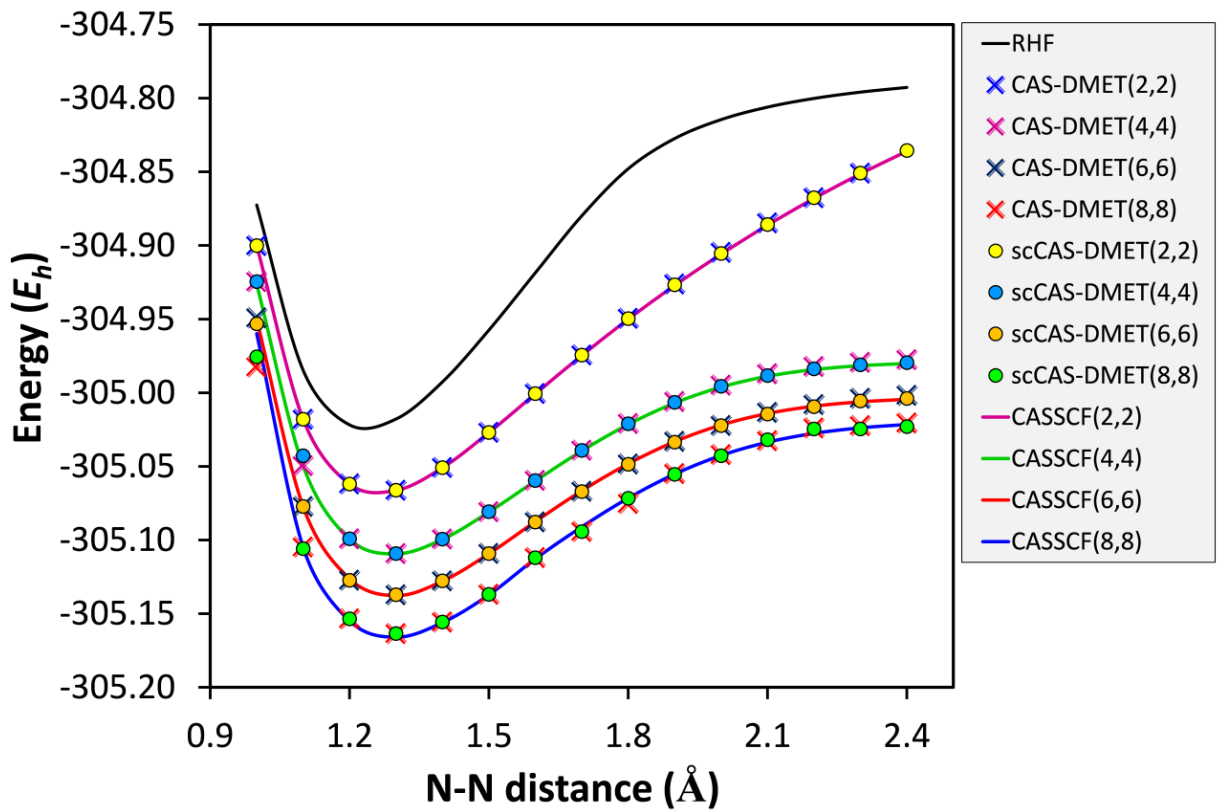

**Figure 9**. Dissociation energy of the N=N bond in pentyldiazene ($CH_3(CH_2)_4$–N=NH).

## V. CONCLUSIONS

We have shown the first example in which CASSCF is used as an impurity solver for DMET. To answer the question posed in our title, overall the performance of CAS-DMET is promising in describing the breaking of an H–H single bond and an N=N double bond with environments of different sizes. The analysis for the embedding and CAS-DMET orbitals rationalizes the difference in energy between CAS-DMET and conventional CASSCF theory. We anticipate that the theory can be improved by incorporating alternative active space partition schemes such as RASSCF or GASSCF followed by multiconfiguration pair density functional theory, to describe both static and dynamic correlation. The combination of these approaches will potentially lead us to generate physically meaningful wave functions for large multireference systems at an affordable cost. Moreover, CAS-DMET can be combined with an automatic active space selection, e.g. the automated selection of active orbital spaces based on DMRG[63] or the atomic valence active space[64] (AVAS) scheme, to improve the efficiency when the interactive optimization of the correlation potential is used. We note that one of the current challenges of DMET is the self-consistent condition that allows to iteratively improve the bath states and the core contribution. The improvement of the convergence rate of both the chemical potential and correlation potential would make the theory more practical in quantum computation on the real chemical systems.

**APPENDIX I: Orbitals transformation in density matrix embedding theory**

There are two primary orbital transformations in a CAS-DMET calculation:

(i) Transforming the canonical orbitals to a set of orthonormal localized orbitals using a unitary matrix $\mathbf{U_1}$,

(ii) Constructing the bath states then transforming the localized orbitals to the embedding space using $\mathbf{U_2} = \sum_{i,j}^{Nf}|\alpha_i\rangle\otimes|\beta_j\rangle$ results in the embedding basis.

The unitary matrix $\mathbf{U_1}$ can be obtained using some localization procedures, such as Löwdin orthogonalizatiton,[65] Boys localization,[66] or intrinsic atomic orbital.[67] The CAS-DMET orbitals are linear combinations of the embedding orbitals (fragment and bath orbitals).

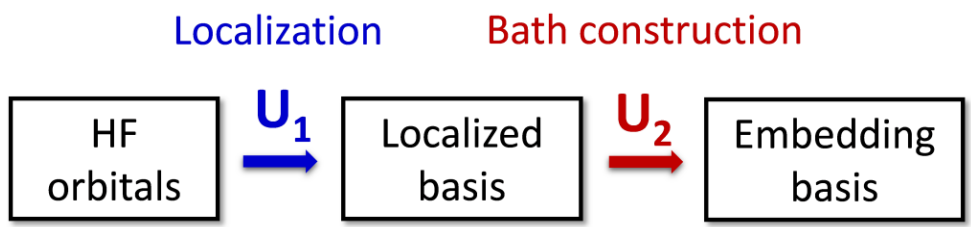

**Figure 10**. Orbital transformation in DMET

The embedding orbitals $\phi_k(r)$ are given by:

$$\phi_k(r) = \sum_{i,k}^{N} \varphi_i(r)(U_1)_{ij}(U_2)_{jk} \tag{A1}$$

where $\varphi_i(r)$ is the finite basis function used in the calculation.

The CAS-DMET orbitals $\Phi_l(r)$ are computed by:

$$\Phi_l(r) = \sum_k^{2N_f} \phi_k(r) C_{kl} \tag{A2}$$

where $C_{kl}$ is the molecular orbital coefficients computed by CASSCF solver. It should be noted that $k$ and $l$ run over at most twice the number of fragments orbitals ($2N_f$).

**ASSOCIATED CONTENT**

**Supporting Information.** Absolute energies are reported with its associated energies. This material is available free of charge via the Internet at http://pubs.acs.org.

**AUTHOR INFORMATION**

**Corresponding Author**

* gagliard@umn.edu (L.G.)

**Author Contributions**

The manuscript was written through contributions of all authors. All authors have given approval to the final version of the manuscript.

Notes

The authors declare no competing financial interests.

ACKNOWLEDGMENT

We thank Gustavo Scuseria, Christopher J. Cramer and Donald G. Truhlar for useful discussion. This work was supported as part of the Inorganometallic Catalyst Design Center, an Energy Frontier Research Center funded by the U.S. Department of Energy, Office of Science, Basic Energy Sciences, under Award DE-SC0012702.

REFERENCES

1. Powell, B. J., Introduction to Effective Low-Energy Hamiltonians in Condensed Matter Physics and Chemistry. In *Computational Methods for Large Systems*, Reimers, J. R., Ed. John Wiley & Sons, Inc.: Hoboken, New Jersey, 2011; pp 309-366.
2. Li, W.; Hua, W.; Fang, T.; Li, S., The Energy-Based Fragmentation Approach for Ab Initio Calculations of Large Systems. In *Computational Methods for Large Systems*, Reimers, J. R., Ed. John Wiley & Sons, Inc.: Hoboken, New Jersey, 2011; pp 225-258.
3. Kowalski, K.; Hammond, J. R.; de Jong, W. A.; Fan, P.-D.; Valiev, M.; Wang, D.; Govind, N., Coupled-Cluster Calculations for Large Molecular and Extended Systems. In *Computational Methods for Large Systems*, Reimers, J. R., Ed. John Wiley & Sons, Inc.: Hoboken, New Jersey, 2011; pp 167-200.
4. Hozoi, L.; Fulde, P., Strongly Correlated Electrons: Renormalized Band Structure Theory and Quantum Chemical Methods. In *Computational Methods for Large Systems*, Reimers, J. R., Ed. John Wiley & Sons, Inc.: Hoboken, New Jersey, 2011; pp 201-224.
5. Clark, T.; Stewart, J. J. P., MNDO-Like Semiempirical Molecular Orbital Theory and Its Application to Large Systems. In *Computational Methods for Large Systems*, Reimers, J. R., Ed. John Wiley & Sons, Inc.: Hoboken, New Jersey, 2011; pp 259-286.
6. Towler, M. D., Quantum Monte Carlo, Or, Solving the Many-Particle Schrödinger Equation Accurately While Retaining Favorable Scaling with System Size. In *Computational Methods for Large Systems*, Reimers, J. R., Ed. John Wiley & Sons, Inc.: Hoboken, New Jersey, 2011; pp 117-166.
7. Hirata, S. Quantum chemistry of macromolecules and solids. *Phys. Chem. Chem. Phys.* **2009**, *11* (38), 8397-8412.
8. Hohenberg, P.; Kohn, W. Inhomogeneous Electron Gas. *Phys. Rev.* **1964**, *136* (3B), B864-B871.

9. Kohn, W.; Sham, L. J. Self-Consistent Equations Including Exchange and Correlation Effects. *Phys. Rev.* **1965**, *140* (4A), A1133-A1138.
10. Cohen, A. J.; Mori-Sánchez, P.; Yang, W. Challenges for Density Functional Theory. *Chem. Rev.* **2012**, *112* (1), 289-320.
11. Stoudenmire, E. M.; Wagner, L. O.; White, S. R.; Burke, K. One-Dimensional Continuum Electronic Structure with the Density-Matrix Renormalization Group and Its Implications for Density-Functional Theory. *Phys. Rev. Lett.* **2012**, *109* (5), 056402.
12. Anisimov, V. I.; Zaanen, J.; Andersen, O. K. Band theory and Mott insulators: Hubbard U instead of Stoner I. *Phys. Rev. B* **1991**, *44* (3), 943-954.
13. Malet, F.; Gori-Giorgi, P. Strong Correlation in Kohn-Sham Density Functional Theory. *Phys. Rev. Lett.* **2012**, *109* (24), 246402.
14. Olsen, J.; Roos, B. O.; Jørgensen, P.; Jensen, H. J. A. Determinant based configuration interaction algorithms for complete and restricted configuration interaction spaces. *J. Chem. Phys.* **1988**, *89* (4), 2185-2192.
15. Knowles, P. J.; Handy, N. C. A new determinant-based full configuration interaction method. *Chem. Phys. Lett.* **1984**, *111* (4), 315-321.
16. Pople, J. A.; Head-Gordon, M.; Raghavachari, K. Quadratic configuration interaction. A general technique for determining electron correlation energies. *J. Chem. Phys.* **1987**, *87* (10), 5968-5975.
17. Ramabhadran, R. O.; Raghavachari, K. Extrapolation to the Gold-Standard in Quantum Chemistry: Computationally Efficient and Accurate CCSD(T) Energies for Large Molecules Using an Automated Thermochemical Hierarchy. *J. Chem. Theory Comput.* **2013**, *9* (9), 3986-3994.
18. Per, S.; Anders, H.; Björn, R.; Bernard, L. A Comparison of the Super-CI and the Newton-Raphson Scheme in the Complete Active Space SCF Method. *Phys. Scr.* **1980**, *21* (3-4), 323.
19. Siegbahn, P. E. M.; Almlöf, J.; Heiberg, A.; Roos, B. O. The complete active space SCF (CASSCF) method in a Newton–Raphson formulation with application to the HNO molecule. *J. Chem. Phys.* **1981**, *74* (4), 2384-2396.
20. Roos, B. O.; Taylor, P. R.; Siegbahn, P. E. M. A complete active space SCF method (CASSCF) using a density matrix formulated super-CI approach. *Chem. Phys.* **1980**, *48* (2), 157-173.
21. Ghosh, S.; Cramer, C. J.; Truhlar, D. G.; Gagliardi, L. Generalized-active-space pair-density functional theory: an efficient method to study large, strongly correlated, conjugated systems. *Chem. Sci.* **2017**, *8* (4), 2741-2750.
22. Vogiatzis, K. D.; Li Manni, G.; Stoneburner, S. J.; Ma, D.; Gagliardi, L. Systematic Expansion of Active Spaces beyond the CASSCF Limit: A GASSCF/SplitGAS Benchmark Study. *J. Chem. Theory Comput.* **2015**, *11* (7), 3010-3021.
23. Ma, D.; Li Manni, G.; Gagliardi, L. The generalized active space concept in multiconfigurational self-consistent field methods. *J. Chem. Phys.* **2011**, *135* (4), 044128.
24. Ivanic, J. Direct configuration interaction and multiconfigurational self-consistent-field method for multiple active spaces with variable occupations. I. Method. *J. Chem. Phys.* **2003**, *119* (18), 9364-9376.
25. Ivanic, J. Direct configuration interaction and multiconfigurational self-consistent-field method for multiple active spaces with variable occupations. II. Application to oxoMn(salen) and N2O4. *J. Chem. Phys.* **2003**, *119* (18), 9377-9385.

26. Andersson, K.; Malmqvist, P. Å.; Roos, B. O. Second-order perturbation theory with a complete active space self-consistent field reference function. *J. Chem. Phys.* **1992**, *96* (2), 1218-1226.
27. Sauri, V.; Serrano-Andrés, L.; Shahi, A. R. M.; Gagliardi, L.; Vancoillie, S.; Pierloot, K. Multiconfigurational Second-Order Perturbation Theory Restricted Active Space (RASPT2) Method for Electronic Excited States: A Benchmark Study. *J. Chem. Theory Comput.* **2011**, *7* (1), 153-168.
28. Malmqvist, P. Å.; Pierloot, K.; Shahi, A. R. M.; Cramer, C. J.; Gagliardi, L. The restricted active space followed by second-order perturbation theory method: Theory and application to the study of CuO2 and Cu2O2 systems. *J. Chem. Phys.* **2008**, *128* (20), 204109.
29. Roskop, L.; Gordon, M. S. Quasi-degenerate second-order perturbation theory for occupation restricted multiple active space self-consistent field reference functions. *J. Chem. Phys.* **2011**, *135* (4), 044101.
30. Li Manni, G.; Carlson, R. K.; Luo, S.; Ma, D.; Olsen, J.; Truhlar, D. G.; Gagliardi, L. Multiconfiguration Pair-Density Functional Theory. *J. Chem. Theory Comput.* **2014**, *10* (9), 3669-3680.
31. Nakatani, N.; Chan, G. K.-L. Efficient tree tensor network states (TTNS) for quantum chemistry: Generalizations of the density matrix renormalization group algorithm. *J. Chem. Phys.* **2013**, *138* (13), 134113.
32. Orús, R. A practical introduction to tensor networks: Matrix product states and projected entangled pair states. *Ann. Phys.* **2014**, *349* (Supplement C), 117-158.
33. Chan, G. K.-L.; Sharma, S. The Density Matrix Renormalization Group in Quantum Chemistry. *Annu. Rev. Phys. Chem.* **2011**, *62* (1), 465-481.
34. Olivares-Amaya, R.; Hu, W.; Nakatani, N.; Sharma, S.; Yang, J.; Chan, G. K.-L. The ab-initio density matrix renormalization group in practice. *J. Chem. Phys.* **2015**, *142* (3), 034102.
35. Chan, G. K.-L.; Head-Gordon, M. Highly correlated calculations with a polynomial cost algorithm: A study of the density matrix renormalization group. *J. Chem. Phys.* **2002**, *116* (11), 4462-4476.
36. Fulde, P.; Stoll, H. Dealing with the exponential wall in electronic structure calculations. *J. Chem. Phys.* **2017**, *146* (19), 194107.
37. Wouters, S.; Jiménez-Hoyos, C. A.; Sun, Q.; Chan, G. K. L. A Practical Guide to Density Matrix Embedding Theory in Quantum Chemistry. *J. Chem. Theory Comput.* **2016**, *12* (6), 2706-2719.
38. Knizia, G.; Chan, G. K.-L. Density Matrix Embedding: A Simple Alternative to Dynamical Mean-Field Theory. *Phys. Rev. Lett.* **2012**, *109* (18), 186404.
39. Zgid, D.; Chan, G. K.-L. Dynamical mean-field theory from a quantum chemical perspective. *J. Chem. Phys.* **2011**, *134* (9), 094115.
40. Kotliar, G.; Savrasov, S. Y.; Haule, K.; Oudovenko, V. S.; Parcollet, O.; Marianetti, C. A. Electronic structure calculations with dynamical mean-field theory. *Rev. Mod. Phys.* **2006**, *78* (3), 865-951.
41. Georges, A.; Kotliar, G.; Krauth, W.; Rozenberg, M. J. Dynamical mean-field theory of strongly correlated fermion systems and the limit of infinite dimensions. *Rev. Mod. Phys.* **1996**, *68* (1), 13-125.
42. Kotliar, G.; Vollhardt, D. Strongly Correlated Materials: Insights From Dynamical Mean-Field Theory. *Phys. Today* **2004**, *57* (3), 53-59.

43. Hirata, S.; Doran, A. E.; Knowles, P. J.; Ortiz, J. V. One-particle many-body Green's function theory: Algebraic recursive definitions, linked-diagram theorem, irreducible-diagram theorem, and general-order algorithms. *J. Chem. Phys.* **2017**, *147* (4), 044108.
44. Bulik, I. W.; Scuseria, G. E.; Dukelsky, J. Density matrix embedding from broken symmetry lattice mean fields. *Phys. Rev. B* **2014**, *89* (3), 035140.
45. Chen, Q.; Booth, G. H.; Sharma, S.; Knizia, G.; Chan, G. K.-L. Intermediate and spin-liquid phase of the half-filled honeycomb Hubbard model. *Phys. Rev. B* **2014**, *89* (16), 165134.
46. Gunst, K.; Wouters, S.; De Baerdemacker, S.; Van Neck, D. Block product density matrix embedding theory for strongly correlated spin systems. *Phys. Rev. B* **2017**, *95* (19), 195127.
47. Fan, Z.; Jie, Q.-l. Cluster density matrix embedding theory for quantum spin systems. *Phys. Rev. B* **2015**, *91* (19), 195118.
48. Knizia, G.; Chan, G. K.-L. Density Matrix Embedding: A Strong-Coupling Quantum Embedding Theory. *J. Chem. Theory Comput.* **2013**, *9* (3), 1428-1432.
49. Bulik, I. W.; Chen, W.; Scuseria, G. E. Electron correlation in solids via density embedding theory. *J. Chem. Phys.* **2014**, *141* (5), 054113.
50. Tsuchimochi, T.; Welborn, M.; Van Voorhis, T. Density matrix embedding in an antisymmetrized geminal power bath. *J. Chem. Phys.* **2015**, *143* (2), 024107.
51. Ricke, N.; Welborn, M.; Ye, H.-Z.; Van Voorhis, T. Performance of Bootstrap Embedding for long-range interactions and 2D systems. *Mol. Phys.* **2017**, *115* (17-18), 2242-2253.
52. Welborn, M.; Tsuchimochi, T.; Van Voorhis, T. Bootstrap embedding: An internally consistent fragment-based method. *J. Chem. Phys.* **2016**, *145* (7), 074102.
53. Zheng, B.-X.; Kretchmer, J. S.; Shi, H.; Zhang, S.; Chan, G. K.-L. Cluster size convergence of the density matrix embedding theory and its dynamical cluster formulation: A study with an auxiliary-field quantum Monte Carlo solver. *Phys. Rev. B* **2017**, *95* (4), 045103.
54. Wouters, S.; A. Jiménez-Hoyos, C.; K.L. Chan, G., Five Years of Density Matrix Embedding Theory. In *Fragmentation*, John Wiley & Sons, Ltd: 2017; pp 227-243.
55. Peschel, I. Special Review: Entanglement in Solvable Many-Particle Models. *Braz. J. Phys.* **2012**, *42* (3), 267-291.
56. Israel, K. Lower entropy bounds and particle number fluctuations in a Fermi sea. *J. Phys. A: Math. Gen.* **2006**, *39* (4), L85.
57. Hosteny, R. P.; Dunning, T. H.; Gilman, R. R.; Pipano, A.; Shavitt, I. Ab initio study of the π-electron states of trans-butadiene. *J. Chem. Phys.* **1975**, *62* (12), 4764-4779.
58. Wouters, S. https://github.com/sebwouters/qc-dmet. **2016**.
59. Pham, H. Q. https://github.com/hungpham2017/casdmet. **2017**.
60. Hariharan, P. C.; Pople, J. A. The influence of polarization functions on molecular orbital hydrogenation energies. *Theor. Chim. Acta.* **1973**, *28* (3), 213-222.
61. Hehre, W. J.; Ditchfield, R.; Pople, J. A. Self—Consistent Molecular Orbital Methods. XII. Further Extensions of Gaussian—Type Basis Sets for Use in Molecular Orbital Studies of Organic Molecules. *J. Chem. Phys.* **1972**, *56* (5), 2257-2261.
62. Sun, Q.; Chan, G. K.-L. Exact and Optimal Quantum Mechanics/Molecular Mechanics Boundaries. *J. Chem. Theory Comput.* **2014**, *10* (9), 3784-3790.
63. Stein, C. J.; Reiher, M. Automated Selection of Active Orbital Spaces. *J. Chem. Theory Comput.* **2016**, *12* (4), 1760-1771.

64. Sayfutyarova, E. R.; Sun, Q.; Chan, G. K.-L.; Knizia, G. Automated Construction of Molecular Active Spaces from Atomic Valence Orbitals. *J. Chem. Theory Comput.* **2017**, *13* (9), 4063-4078.
65. Löwdin, P. O. On the Non-Orthogonality Problem Connected with the Use of Atomic Wave Functions in the Theory of Molecules and Crystals. *J. Chem. Phys.* **1950**, *18* (3), 365-375.
66. Foster, J. M.; Boys, S. F. Canonical Configurational Interaction Procedure. *Rev. Mod. Phys.* **1960**, *32* (2), 300-302.
67. Knizia, G. Intrinsic Atomic Orbitals: An Unbiased Bridge between Quantum Theory and Chemical Concepts. *J. Chem. Theory Comput.* **2013**, *9* (11), 4834-4843.

**Table-of-Contents Graphic**

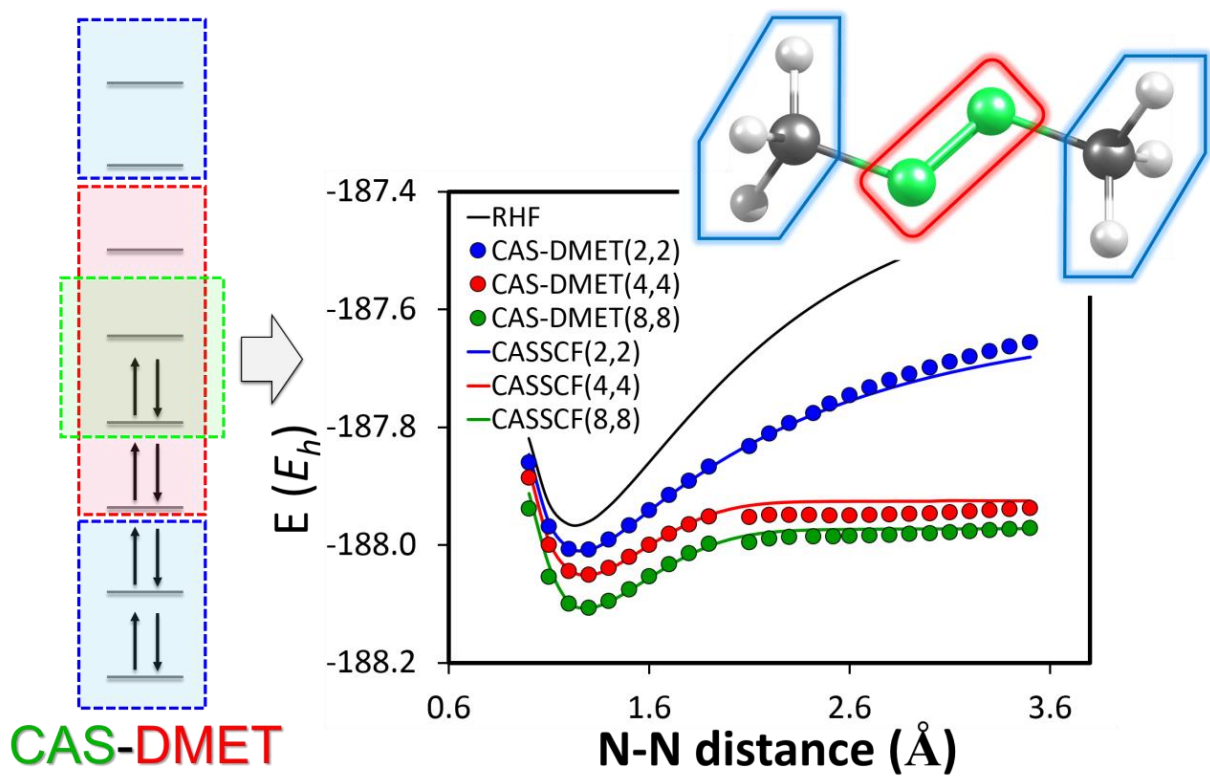